\documentclass[reprint,superscriptaddress,amsmath,amssymb, aps, prx]{revtex4-2}

\usepackage{graphicx}
\usepackage{hyperref}
\usepackage{booktabs}
\usepackage{siunitx}[=v2]
\usepackage[utf8]{inputenc}

\begin{document}

\title{A strontium quantum-gas microscope}

\author{Sandra Buob}%
\affiliation{ICFO - Institut de Ciencies Fotoniques, The Barcelona Institute of Science and Technology, 08860 Castelldefels (Barcelona), Spain}%
\author{Jonatan H\"oschele}%
\affiliation{ICFO - Institut de Ciencies Fotoniques, The Barcelona Institute of Science and Technology, 08860 Castelldefels (Barcelona), Spain}%
\author{Vasiliy Makhalov}%
\affiliation{ICFO - Institut de Ciencies Fotoniques, The Barcelona Institute of Science and Technology, 08860 Castelldefels (Barcelona), Spain}%
\author{\\Antonio Rubio-Abadal}%
 \email{antonio.rubio@icfo.eu}%
\affiliation{ICFO - Institut de Ciencies Fotoniques, The Barcelona Institute of Science and Technology, 08860 Castelldefels (Barcelona), Spain}%
\author{Leticia Tarruell}%
 \email{leticia.tarruell@icfo.eu}%
\affiliation{ICFO - Institut de Ciencies Fotoniques, The Barcelona Institute of Science and Technology, 08860 Castelldefels (Barcelona), Spain}%
\affiliation{ICREA, Pg. Llu\'{i}s Companys 23, 08010 Barcelona, Spain}

\date{\today}

\begin{abstract}
The development of quantum-gas microscopes has brought novel ways of probing quantum degenerate many-body systems at the single-atom level. Until now, most of these setups have focused on alkali atoms. Expanding quantum-gas microscopy to alkaline-earth elements will provide new tools, such as SU($N$)-symmetric fermionic isotopes or ultranarrow optical transitions, to the field of quantum simulation. Here, we demonstrate the site-resolved imaging of a $^{84}$Sr bosonic quantum gas in a Hubbard-regime optical lattice. The quantum gas is confined by a two-dimensional in-plane lattice and a light-sheet potential, which operate at the strontium clock-magic wavelength of \SI{813.4}{nm}. We realize fluorescence imaging using the broad \SI{461}{nm} transition, which provides high spatial resolution. Simultaneously, we perform attractive Sisyphus cooling with the narrow \SI{689}{nm} intercombination line. We reconstruct the atomic occupation from the fluorescence images, obtaining imaging fidelities above $94\%$. Finally, we realize a $^{84}$Sr superfluid in the Bose-Hubbard regime. We observe its interference pattern upon expansion, a probe of phase coherence, with single-atom resolution. Our strontium quantum-gas microscope provides a new platform to study dissipative Hubbard models, quantum optics in atomic arrays, and SU($N$) fermions at the microscopic level.
\end{abstract}
\maketitle

\section{\label{sec:introduction}Introduction}

The ability to control and detect individual particles has been a major driving force in the field of quantum science and technology. A prime example is given by quantum-gas microscopes~\cite{gross_quantum_2021}, which are an extremely versatile tool for the analog quantum simulation of Hubbard models. These are devices that provide access to the microscopic properties of quantum gases in optical lattices through site-resolved detection of single atoms. Quantum-gas microscopy, demonstrated for both bosons~\cite{bakr_probing_2010,sherson_singleatomresolved_2010} and fermions~\cite{haller_singleatom_2015, cheuk_quantumgas_2015, parsons_siteresolved_2015, omran_microscopic_2015}, has provided a microscopic point of view on fundamental problems of quantum magnetism~\cite{simon_quantum_2011,fukuhara_quantum_2013, mazurenko_coldatom_2017, wei_quantum_2022}, transport properties in many-body systems~\cite{brown_bad_2019, nichols_spin_2019, guardado-sanchez_subdiffusion_2020} or quantum thermalization~\cite{kaufman_quantum_2016, choi_exploring_2016, rispoli_quantum_2019}, among other examples. Most microscope setups to date have been realized with alkali atoms, which have only one valence electron. Extending quantum-gas microscopy to other atomic species opens exciting opportunities for quantum simulation. So far, examples of beyond-alkali microscopes include bosonic ytterbium~\cite{miranda_siteresolved_2015, yamamoto_ytterbium_2016, miranda_siteresolved_2017}, and highly magnetic erbium~\cite{su_dipolar_2023a}.

Atomic species with two valence electrons, such as alkaline-earth and alkaline-earth-like atoms, are of special interest for quantum simulation~\cite{schafer_tools_2020}.  Their distinct electronic structure, with singlet and triplet states, gives rise to narrow and ultranarrow optical transitions, which can be used as a precise probe, and to long-lived metastable states in which quantum information can be stored~\cite{daley_quantum_2008}. Furthermore, fermionic isotopes with a nuclear spin display inter-particle interactions that are independent of their magnetic state, hence allowing to study SU($N$)-symmetric systems~\cite{cazalilla_ultracold_2014}.
Many of these properties have been exploited in recent optical-lattice works that relied on global observables. Some examples include spin-orbit coupling using the metastable clock state~\cite{livi_synthetic_2016, kolkowitz_spin_2017}, dissipative Bose-Hubbard models~\cite{tomita_observation_2017, bouganne_anomalous_2020}, or SU($N$) fermions~\cite{taie_su_2012, hofrichter_direct_2016, campbell_fermidegenerate_2017, taie_observation_2022}.

Atomic strontium has become a popular choice for the aforementioned research fields. It displays both bosonic and fermionic isotopes which have been successfully brought to quantum degeneracy~\cite{stellmer_boseeinstein_2009, deescobar_boseeinstein_2009, desalvo_degenerate_2010, stellmer_production_2013}. The fermionic isotope $^{87}\text{Sr}$ features a large nuclear spin, leading to a SU($N = 10$) symmetry. 
Furthermore, the ultra-narrow transition at \SI{698}{nm} finds application in optical atomic clocks~\cite{ludlow_optical_2015}, which rely on state-insensitive traps at magic wavelengths~\cite{ye_quantum_2008}.
This set of favorable properties for quantum simulation makes the development of a strontium quantum-gas microscope highly desirable.

Single-atom imaging of strontium atoms has been recently achieved in optical tweezers~\cite{norcia_microscopic_2018, cooper_alkalineearth_2018}, by exploiting the short wavelength of its blue \SI{461}{nm} transition and the cooling properties of its red \SI{689}{nm} intercombination line.
Additionally, hybrid lattice-tweezer platforms have been developed, which have demonstrated single-atom imaging of strontium in optical lattices~\cite{schine_longlived_2022, tao_highfidelity_} and realized programmable quantum walks~\cite{young_tweezerprogrammable_2022,young_atomic_}. However, these studies of hybrid systems have so far focused on the $^{88}\mathrm{Sr}$ isotope which, due to its negligible scattering length, precludes attaining the interacting Hubbard regime or reaching quantum degeneracy through evaporative cooling. 

\begin{figure*}
\includegraphics[width= 1\textwidth]{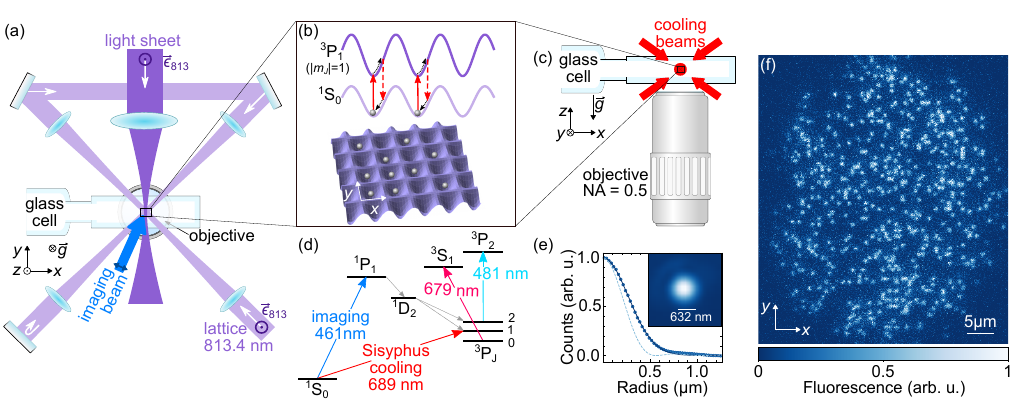}
\caption{Strontium quantum-gas microscope. (a) Top view of the experimental setup, showing the lattice beam passing four times through the quartz glass cell. The light-sheet beam is tightly focused (waist in $z$-direction $\sim$\SI{3.5}{\micro m}) and provides the vertical confinement into the plane. Both lattice and light-sheet beams are vertically polarized ($\vec{\epsilon}_{813}$) and operate at \SI{813.4}{nm} wavelength. The objective, placed below the glass cell, captures the fluorescence signal of the atoms generated by a blue imaging beam at \SI{461}{nm} which is horizontally polarized. (b) Two-dimensional potential created by the four-fold interference of the lattice beams (bottom). The tighter confinement in the $^3$P$_1$ excited state compared to the $^1$S$_0$ ground state (top) allows for attractive Sisyphus cooling on the red intercombination transition at \SI{689}{nm}. (c) Side view of the experimental setup, depicting the six red cooling beams and the objective placed below the glass cell. (d) Energy level diagram of strontium, with the broad-linewidth imaging transition and the narrow-linewidth cooling transition. Shown are also the repumping transitions (\SI{481}{nm}~\cite{hu_analyzing_2019} and \SI{679}{nm}~\cite{dinneen_cold_1999}) used to recover atoms leaked into the long-lived states $^3$P$_{J=0,2}$. (e) Azimuthal average of the point spread function (PSF) (blue circles), with a Gaussian fit (solid line) and the ideal Airy disk (dashed line).  Inset: PSF obtained by averaging over 1187 isolated atoms. (f) Raw fluorescence image of a dilute thermal cloud showing around $500$ individual \textsuperscript{84}Sr atoms.}
\label{fig:fig1}
\end{figure*}

In this work, we demonstrate the realization of a strontium quantum-gas microscope. We achieve this by combining the preparation of a quantum degenerate cloud of \textsuperscript{84}Sr, an interacting bosonic isotope, with the imaging techniques pioneered by the optical-tweezer community. An evaporatively cooled \textsuperscript{84}Sr cloud is confined in a 2D geometry by a highly anisotropic light-sheet optical trap, and loaded into an in-plane square optical lattice, all operating at \SI{813.4}{nm}. This wavelength is magic for the strontium clock transition~\cite{ye_quantum_2008}. For the imaging stage, the confinement is brought to its maximum experimentally available depth and then fluorescence photons, scattered at the broad-linewidth blue transition, are collected with a high numerical aperture objective. The atoms are simultaneously cooled within their lattice sites by using the narrow-linewidth red transition. We realize this in a configuration with a tighter confinement for the $^{3}\mathrm{P}_1$ excited state than for the $^{1}\mathrm{S}_0$ ground state. This difference leads to the so-called attractive Sisyphus cooling~\cite{covey_2000times_2019}, which we demonstrate here in an optical lattice. The high efficiency of this cooling approach highlights that quantum-gas microscopy can be realized with moderate laser powers and in a setup with reduced experimental complexity. Finally, by preparing a quantum gas in a shallow optical lattice, we realize a superfluid in the Bose-Hubbard regime, and detect the interference pattern arising during its expansion due to phase coherence.

We organize this article as follows: In Sec.~\ref{sec:sec2} we describe the main elements of the experimental setup, and the techniques required for single-atom imaging. In Sec.~\ref{sec:sec3} we present a reconstruction of raw fluorescence images and benchmark the site-resolved imaging fidelity of the experiment. Section~\ref{sec:sec4} discusses the spatial dependence of Sisyphus cooling, and how the overall confinement in the system can be characterized from it. In Sec.~\ref{sec:sec5} we discuss the loading of a degenerate gas into the optical lattice and show its interference pattern after expansion, which showcases the phase coherence of a lattice superfluid. Finally, we conclude by summarizing the main results and possible future directions in Sec.~\ref{sec:conclusion}.

\section{\label{sec:sec2}Experimental setup}

The experiment takes place in a glass cell with a high-resolution objective placed right below, which constitute the central pieces of the experimental setup, see Fig.~\ref{fig:fig1} (a)-(c). Inside the glass cell, we trap and cool $^{84}$Sr atoms to quantum degeneracy before loading them into an optical lattice and imaging them with single-site resolution.

\subsection{Preparation of a quantum gas}

For cooling and trapping, we apply the common two-stage magneto-optical trap (MOT) scheme of strontium experiments (see App.~\ref{sec:appendixA} and our previous work~\cite{hoschele_atomnumber_2023}). It results in a cloud of $5\times 10^5\,$atoms with temperatures of few microkelvin. This cold cloud is subsequently loaded into a crossed optical dipole trap, consisting of two \SI{1064}{nm} laser beams at a power of \SI{3}{W} each. This results in a potential depth of \SI{20}{\micro K} where we initially trap $10^5\,$atoms. While holding them in the crossed trap we ramp up a light-sheet potential, which is a tightly focused elliptical beam at the clock-magic wavelength of strontium \SI{813.4}{nm} (see Fig.~\ref{fig:fig1} (a)). The light-sheet-beam waists of roughly \SI{3.5}{\micro m} $\times$ \SI{60}{\micro m} result in a trap depth of \SI{50}{\micro K} already at a power of \SI{270}{mW}.
Taking advantage of the favorable scattering length of \textsuperscript{84}Sr, $a_{s}=123\,a_0$~\cite{stellmer_boseeinstein_2009, deescobar_boseeinstein_2009}, we perform evaporative cooling in this combined potential for \SI{8}{s}. To this end, we exponentially decrease the power of the cross trap beams to zero and the light sheet potential to a finite depth at which we reach quantum degeneracy.

\subsection{Square optical-lattice potential}

We generate the lattice potential with a $\lambda = \SI{813.4}{nm}$ vertically polarized beam, which travels in a bow-tie configuration twice through the glass cell before being retro-reflected, see Fig.~\ref{fig:fig1} (a). This configuration results in the interference of four beams, which all share the same polarization $\vec{\epsilon}_{813}$. The interference leads to a square lattice with spacing $\lambda/\sqrt{2} \approx \SI{575}{nm}$~\cite{sebby-strabley_lattice_2006}, as depicted in Fig.~\ref{fig:fig1} (b), which we use both as a physics and imaging lattice. Additionally, it makes phase stabilization redundant and allows us to recycle the power of the beam. This is highly beneficial for high-depth applications during imaging in quantum-gas microscopes~\cite{brown_spinimbalance_2017, yang_siteresolved_2021, kwon_siteresolved_2022}. 
We achieve a full lattice depth of up to \SI{140}{\micro K} with a total lattice power of \SI{3}{W}. This corresponds to 800\,$E_r$, where $E_r = h^2/2 m \lambda^2 \approx \, h \times \SI{3.6}{kHz}$ is the recoil energy of the lattice beam photons. Here $h$ is the Planck's constant and $m$ is the mass of $^{84}$Sr.

\subsection{In-lattice Sisyphus cooling}
An efficient cooling mechanism is required to preserve the atoms in their individual lattice sites during imaging. For atomic strontium, the narrow linewidth of the red transition can be exploited for resolved sideband cooling, which requires equal trapping frequencies for ground and excited states ($\omega_g = \omega_e$) \cite{ido_recoilfree_2003, ye_quantum_2008}. Imaging of individual strontium atoms has been shown with this technique by tuning to a magic configuration in both optical tweezers and lattices~\cite{cooper_alkalineearth_2018, norcia_microscopic_2018, schine_longlived_2022}. A different approach, available in the case of nonzero differential trapping, is Sisyphus cooling~\cite{taieb_cooling_1994, ivanov_laserdriven_2011}. Depending on the sign of the differential trapping, this method can be realized in either the repulsive ($\omega_g > \omega_e$) or attractive ($\omega_g < \omega_e$) regime. Single-atom imaging in the repulsive Sisyphus regime has been demonstrated in optical tweezers~\cite{cooper_alkalineearth_2018} and lattices~\cite{tao_highfidelity_}, while in the attractive regime it has only been shown in optical tweezers~\cite{cooper_alkalineearth_2018, covey_2000times_2019}.

In our experiment, we perform attractive Sisyphus cooling in an optical lattice. As depicted in Fig.~\ref{fig:fig1} (b), the excited states with $|m_{J}| =1$ experience a deeper trapping potential than the ground state. This is the case when a magnetic bias field is parallel to the lattice and light-sheet polarization, i.e., $\vec{B} \parallel \vec{\epsilon}_{813}$. As atoms are excited from the bottom of the trap via the \SI{689}{nm} transition, they will experience an average reduction of energy through spontaneous emission, due to the steeper potential of the excited state. While Sisyphus cooling can be realized with a single laser beam~\cite{covey_2000times_2019}, here we use the same set of laser beams as for the red MOT stage, see Fig.~\ref{fig:fig1} (c), reducing the complexity of the experimental setup. The total intensity of the beams during imaging is $\sim1000\, I_{s,\mathrm{red}}$, where  $I_{s,\mathrm{red}}$ is the saturation intensity of the red transition. The frequency is set to the light-shifted resonance, which in the center of the cloud is up to \SI{1.3}{MHz} red detuned from its free-space value (see App. \ref{sec:appendixC}). Due to light-assisted collisions, the cooling process also induces parity projection to the occupation in each lattice site~\cite{cooper_alkalineearth_2018}.

During the imaging stage, both lattice and light sheet are brought to its largest depth, with on-site trapping frequencies of  $(\omega_{x}, \omega_{y}, \omega_{z})_\mathrm{site}\approx 2\pi\times$(150, 140, 6.5)\,kHz. Efficient in-trap cooling typically requires Lamb-Dicke parameters $\eta_{i} = \sqrt{ h \pi  /m \omega_{i} \lambda_{c}^2}<1$ for its three axes $i =x,y,z$, where $\lambda_c$ is the wavelength of the cooling transition. In our case, for $\lambda_c = \SI{689}{nm}$, we obtain a small in-plane parameter $\eta_{x, y} \approx 0.2$ and an axial parameter $\eta_{z} \approx 0.9$. We observe efficient cooling even with such a weak axial confinement, which is consistent with a recent result performing attractive Sisyphus cooling in shallow optical tweezers~\cite{urech_narrowline_2022}.

\subsection{High-resolution fluorescence imaging}

The fluorescence imaging of the atoms is realized by scattering photons on the broad-linewidth \SI{461}{nm} transition, through a near-resonant beam shone into the atomic cloud. The photons are collected by a high-resolution infinity-corrected objective (G Plan Apo 50X, Mitutoyo) with a numerical aperture (NA) of $0.5$. The objective is placed in working distance (\SI{15.08}{mm}) from the atoms, outside a quartz glass cell with \SI{3.5}{mm} thickness (see Fig.~\ref{fig:fig1} (c)). The imaging system consists additionally of an achromatic doublet (ACT508-500-A, Thorlabs) which leads to a measured magnification of $126.7(3)$. The fluorescence photons are detected by an EMCCD camera (Andor iXon Ultra 897) with \SI{85}{\%} quantum efficiency at \SI{461}{nm} and a $\SI{16}{\micro m}$ pixel size. 

To induce fluorescence, we shine a linearly polarized blue laser with $I\sim 10^{-3}I_{s,\mathrm{blue}}$, detuned from resonance by $\delta \approx -2 \, \Gamma_\mathrm{blue}$. Here $I_{s,\mathrm{blue}}$ and $\Gamma_\mathrm{blue}$ are respectively the saturation intensity and the linewidth of the blue transition. The linear polarization is chosen to be in plane, which ensures optimal collection efficiency of the radiation pattern~\cite{cooper_alkalineearth_2018}. From the intensity and detuning we estimate a scattering rate of $\sim \,$\SI{3}{kHz}, which takes into account the simultaneous saturation of the red cooling transition. The imaging system has an estimated collection efficiency of \SI{6.7}{\%}, (\SI{71}{\%} transmission of the path, and an objective collection efficiency of \SI{9.4}{\%}). In our pictures, we typically detect around $300$ photons per atom during a 3-second exposure. In Fig.~\ref{fig:fig1} (f), we show a fluorescence image of a thermal cloud with around $500$ atoms. In addition to the blue beam, we simultaneously shine two repumping beams to recover atoms leaked into the long-lived $^{3}\mathrm{P}_2$ and $^{3}\mathrm{P}_0$ states, see Fig.~\ref{fig:fig1} (d).

Based on the numerical aperture of the objective, we estimate a Rayleigh resolution of \SI{562}{nm}, which corresponds to a full width at half maximum (FWHM) of 470 nm. Such a resolution is below the \SI{575}{nm} lattice spacing of the four-fold optical lattice, and can hence resolve the atoms in the individual sites. In order to characterize our imaging system, we measure its point spread function (PSF), see Fig.~\ref{fig:fig1} (e). For this purpose, we capture the fluorescence signal of isolated atoms in sparsely filled lattices, overlap and average their signal. From the resulting PSF, we find a FWHM of \SI{632(3)}{nm}. This value is above the ideal expectation, but it is sufficiently small to resolve individual lattice sites in the system.

\begin{figure}
\includegraphics[width= \columnwidth]{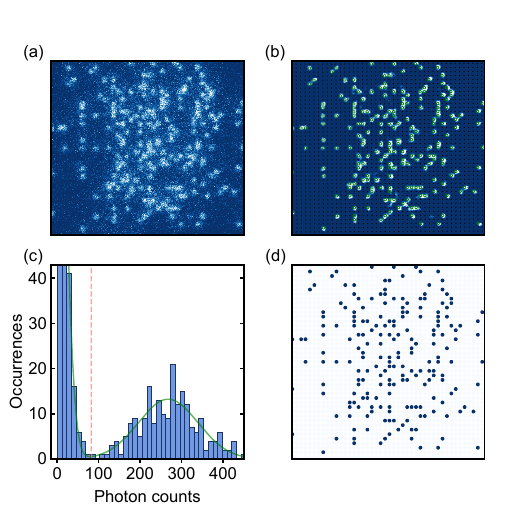}
\caption{Reconstruction of the atomic occupation from a fluorescence image. (a) Raw experimental image of a thermal cloud loaded in the lattice. (b) Deconvolution of the image in (a) with the PSF to obtain a cleaner signal. The black circles mark the positions of the reconstructed lattice grid, while the green circles indicate the lattice sites detected as occupied. (c) Histogram of the counts at each lattice site for the deconvolved picture at (b). The plot indicates the distinguishability between the zero-atom (left peak) and one-atom (right peak) cases. To choose the threshold value in counts, we fit a double Gaussian function to the histogram. From the fit we estimate that $>99.9\%$ of the one-atom distribution lies above the threshold in this image. (d) Binary occupation obtained from the reconstruction of the picture. }
\label{fig:fig2}
\end{figure}

\section{\label{sec:sec3} Image reconstruction and pinning fidelity}

In order to access the microscopic properties of the cloud, one has to reconstruct the atomic occupation from the fluorescence images and hence obtain the site-resolved distribution, see Fig.~\ref{fig:fig2}. As a first step, the captured images are deconvolved with the measured PSF of the system by realizing 15 iterations of the Richardson-Lucy algorithm~\cite{richardson_bayesianbased_1972, lucy_iterative_1974}. The resulting image, shown in Fig.~\ref{fig:fig2} (b), is Fourier transformed in order to extract the lattice vectors and their phase, which enable a full reconstruction of the lattice grid. From the lattice reconstruction, we find an anisotropy in the lattice spacing, with $a_x=\SI{608(1)}{nm}$ and $a_y=\SI{547(1)}{nm}$. It can be explained by a deviation of the interfering beams from the ideal $90^{\circ}$ angles, see App. \ref{sec:appendixA}. With the lattice reconstructed, we bin the photon counts at each lattice site to obtain the histogram in Fig.~\ref{fig:fig2} (c). By setting a threshold in the number of binned photon counts, we are able to discriminate occupied lattice sites from empty ones, which we display in Fig.~\ref{fig:fig2} (d). The separation between the zero-photon peak and the right peak, which is centered at around 300 photon counts, indicates a reliable detection of single-occupied sites. 

\begin{figure}
\includegraphics[width= \columnwidth]{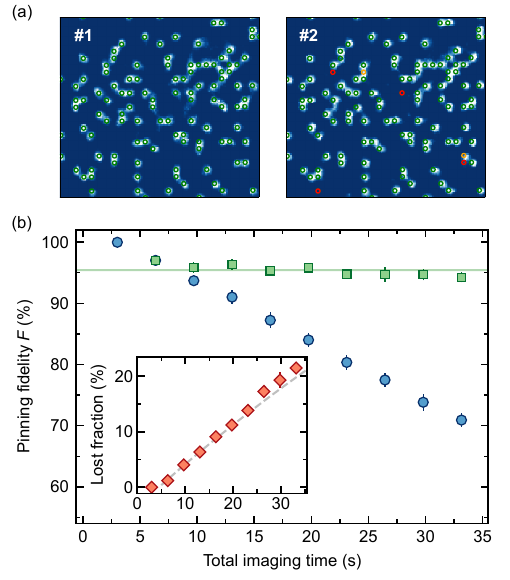}
\caption{Fidelity of the site-resolved single-atom detection. (a) Two consecutive images ($\SI{15.5}{\micro m} \times \SI{15.5}{\micro m}$ region) of the same cloud after being deconvolved. In the first picture, we mark sites where an atom has been detected with green circles. In the second picture, lost atoms are indicated with red circles, atoms that hopped in orange, and those that remained in the same position in green. The single-shot pinning fidelity is $F = 96.5\%$. (b) Pinning fidelity $F$ as a function of imaging time. We take 10 images of the same cloud within \SI{35}{s} and compare the reconstructed occupation to either the initial shot (blue circles) or the previous shot (green squares). The solid green line marks the average of all green-square fidelity data, which is around $95\%$. The inset shows the accumulated losses for each consecutive measurement (red diamonds) compared to those caused by background-gas losses (dashed gray line). Error bars indicate one standard error of the mean. 
}
\label{fig:fig3}
\end{figure}

To benchmark the imaging fidelity of our system, we measure a set of pairs of consecutive images of the same atomic cloud, applying the reconstruction algorithm to both images. This allows us to identify loss and hopping events from one shot to the next one, and to estimate the pinning fidelity, i.e., the fraction of atoms that remain in their original sites in the second image. In Fig.~\ref{fig:fig3} (a) we show the deconvolutions of two exemplary consecutive images. The green circles in the first picture (left) indicate the sites where an atom has been detected. In the second picture (right), green circles indicate atoms identified in both pictures, red circles atoms that have left their original site and orange circles newly identified atoms, i.e., hopping events. From averaging over 47 such pairs of images ($10\% $ average density over a  25$\times$\SI{25}{\micro m} region) we extract a pinning fidelity of $94.1(2) \%$, a hopping fraction of $1.54(14) \%$, and a loss fraction of $4.3(3) \%$.

To further characterize the cooling efficiency and pinning fidelity of the system, we take a set of 10 consecutive images for the same realization, similarly as in~\cite{kwon_siteresolved_2022}. In Fig.~\ref{fig:fig3} (b), we show the pinning fidelity with respect to the first picture (blue circles), which after \SI{35}{s} of imaging has decreased to $\sim 70\%$. We also extract the relative fidelity between consecutive images (green squares), which is shown to stay around $95\%$ over the whole imaging time. We use this measurement to extract the losses as a function of imaging time, as shown in the inset (red diamonds). These losses are almost entirely explained by those associated to background-gas collisions, measured in the lattice without blue imaging light (gray dashed line) and characterized by an exponential decay constant $\tau_\mathrm{loss}$ =130(5)\,s. This is evidence that, at the chosen imaging parameters, losses associated to the imaging process are essentially negligible.

\section{\label{sec:sec4}Spatial dependence of Sisyphus cooling}

Sisyphus cooling displays an intrinsic dependence of the cooling resonance on the trap intensities, due to the finite differential AC Stark shift between ground and excited states. This implies that the cooling performance is sensitive to the overall shape of the system confinement, and hence can be used as a tool to characterize it.
In the configuration of our setup, with $\vec{B} \parallel \vec{\epsilon}_{813}$, the ground and excited state polarizabilities are given by $\alpha_g = 286.0$ a.u. and $\alpha_e = 353.9$ a.u.~\cite{madjarov_entangling_2021}, which corresponds to $\omega_{g}/\omega_{e}=\sqrt{\alpha_g/\alpha_e} \approx 0.9$. To characterize the spatial inhomogeneity of the system, here we study the cloud sizes of fluorescence images for different frequencies of the cooling laser. 

\begin{figure}
\includegraphics[width= \columnwidth]{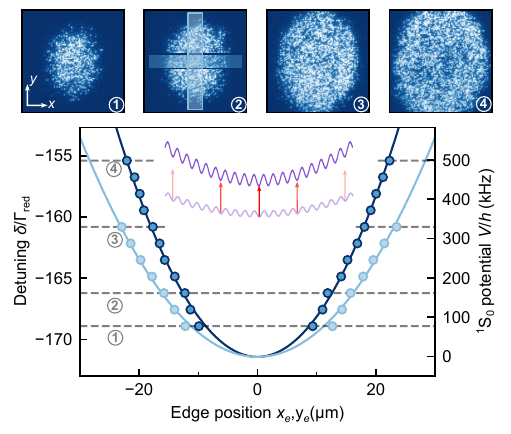}
\caption{Spatial inhomogeneity of Sisyphus cooling. Top row: Four individual fluorescence images of atomic clouds at different red-cooling detunings. Frequencies closer to the free-space resonance lead to larger cloud sizes. Bottom: red cooling frequency corresponding to the position of the cloud edge. The edge positions of the cloud along $x$ and $y$ axes are extracted by fitting an error function to the fluorescence signal within a sliced region (see rectangular boxes in the second fluorescence image) integrated along the perpendicular direction. The shape allows us to extract the trapping frequency directly from the spectral dependence of the cloud size. The right axis indicates the potential $V$ experienced by the ground state $^{1}\mathrm{S}_0$, computed from the differential polarizability. The inset in the lower plot illustrates the overall differential trapping, and how it shifts the red transition out of resonance. Horizontal dashed lines indicate the frequencies corresponding to the pictures of the top row. The error bars are smaller than the marker size.}
\label{fig:fig4}
\end{figure}

In Fig.~\ref{fig:fig4}, we show four fluorescence images of large thermal clouds for increasing frequencies of the cooling laser. The first image displays a small elliptical cloud, explained by efficient cooling in the center and a complete atom loss in its outer edges. As the cooling frequency is increased, the red light gets closer to resonance with the atoms in the boundary. These atoms in turn experience a smaller light shift, and the fluorescent cloud size increases. Eventually, for even higher frequencies, the center of the cloud is no longer efficiently cooled, and losses lead to a density reduction in that region. In practice, we select a red-laser frequency that ensures efficient cooling, as indicated by the local pinning fidelities $F$, over the entire size of the prepared system. This is obtained at the expense of a reduced field of view in the images. 

The spectral dependence of the cloud size can be used to spatially characterize the overall potential experienced by the atoms. To do so, we extract for each picture the positions of the two edges of the cloud along $x$ and $y$ directions, which constitute equipotential points in the cloud. We realize this by selecting a region of interest for each direction (blue rectangular boxes in Fig.~\ref{fig:fig4}), integrating the fluorescence along the perpendicular direction, and fitting error functions to the resulting one-dimensional data. The fit parameters provide us the edge values $x_e$,\, $y_e$ for each cooling frequency. We present the results in Fig.~\ref{fig:fig4}, in which we plot the set cooling detunings as a function of the extracted edge positions. There, we identify the symmetric distribution of the frequencies around the center of the cloud, consistent with the light shift of a harmonic potential. 
Additionally, we infer the spatial potential experienced by the ground state $^{1}\mathrm{S}_0$, represented in the right axis, by using the known polarizabilities of the involved states.
By performing a combined fit of the harmonic potentials for both directions, which share the same potential offset, we can extract the trapping frequencies of the overall potential $(\omega_{x}, \omega_{y})_\mathrm{image} = 2 \pi \times  (497(2),387(2))\,$Hz. The measured anisotropy in the trap frequencies, as indicated by the elliptical shape of the fluorescent clouds, originates from the elliptical confinement of the light sheet in the plane, which we measured independently via dipole oscillations to be $(\omega_{x}, \omega_{y})_\mathrm{sheet} = 2\pi \times (420(10), 252(5))\,$Hz. The overall confinement, as characterized in Fig.~\ref{fig:fig4}, is compatible with the combination of the potentials from the light sheet and the optical-lattice envelope.

\section{\label{sec:sec5}Bose-Hubbard superfluid}
Finally, we exploit our quantum-gas microscope to perform single-site imaging of a \textsuperscript{84}Sr superfluid in the Hubbard regime. While the images discussed in the previous sections were obtained by loading a thermal cloud into the lattice, here we perform evaporative cooling for $t_\mathrm{evap} = \SI{8}{s}$ and load the resulting Bose-Einstein condensate adiabatically into a shallow optical lattice. For lattice depths $V_0 \gtrsim 2 E_r$, we expect the tight-binding approximation to hold, and the system to be described by the 2D Bose-Hubbard model with a tunneling amplitude $J$ and interaction strength $U$. For $J/U > 0.06$~\cite{capogrosso-sansone_monte_2008} the cloud should be in a superfluid phase with phase coherence across the entire system. This is the case in our setup for depths  $V_0 = 2-5 \, E_r$ due to the relatively weak vertical confinement of $\sim\SI{1.1(1)}{kHz}$ used in this section. As before, $E_r$ is defined in terms of the single-photon recoil and not of the lattice spacing, see App.~\ref{sec:appendixB}.

\begin{figure}[ht]
\includegraphics[width= \columnwidth]{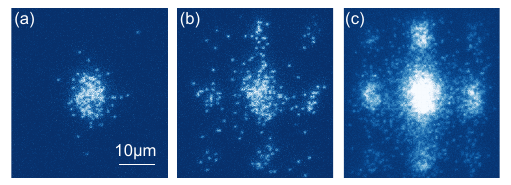}
\caption{Detection of a Bose-Hubbard superfluid. (a) In situ image of a quantum gas loaded into the optical lattice at a depth of $V_0 = 2.3(1)\,E_r$. (b) Image taken after an in-trap 2D expansion, during which the lattice is abruptly switched off for $t_{\text{exp}}=\SI{2}{ms}$ while holding the atoms in plane by means of the harmonic trap produced by the light sheet beam. An average over 10 such pictures is displayed in (c), where higher order interference peaks are visible at the corners of the image. The color scale in (c) has been adjusted to better discern the pattern.}
\label{fig:fig5}
\end{figure}

Fig.~\ref{fig:fig5} (a) shows the in situ image of an evaporatively cooled cloud loaded into a lattice of depth $V_0 = 2.3(1) E_r$, for which we expect a superfluid state. For a lattice potential with isotropic spacing and depth, this corresponds to a tunneling $J/h=\SI{188(8)}{Hz}$ and an interaction strength of $U/h=\SI{121(3)}{Hz}$. In our system, the lattice-spacing anisotropy and the relative attenuation of the lattice beams leads to an asymmetry in the tunneling amplitudes. We estimate a value $J_x/J_y \approx 0.7$, see App.~\ref{sec:appendixB}, for which the system remains in the superfluid Hubbard regime.

An atomic  superfluid in an optical lattice is commonly demonstrated by the observation of an interference pattern after a free-space time-of-flight~\cite{greiner_quantum_2002}. Here, we perform instead an in-plane expansion of the cloud within a shallow light sheet potential after abruptly switching off the lattice potential. After the expansion, a sudden ramp-up of the lattice pins the atoms at their position, followed by site-resolved fluorescence imaging. In Fig.~\ref{fig:fig5} (b), we show a fluorescence image taken after an expansion of $t_{\text{exp}}=\SI{2}{ms}$, where we observe the interference pattern.
An important advantage of this method is that it enables the detection of phase coherence in the system even with low atom numbers (typically a few hundreds), as previously shown in other quantum-gas experiments~\cite{bakr_probing_2010, ozawa_observation_2023a}. In our system, the anisotropy in the light-sheet confinement, with frequencies $ (\omega_{x}, \omega_{y})_\mathrm{exp} =  2\pi \times (62.0(7) , 38.8(8)$)\,Hz, leads to an asymmetric expansion in the two in-plane directions. The chosen expansion time $t_{\text{exp}}=\SI{2}{ms}$ corresponds to $T_{x}/8 = (2 \pi/\omega_{x} )/ 8$. While mapping of the momentum distribution into the density distribution requires $\sim T/4$, we deliberately chose $T_{x}/8$ to avoid high-density peaks which would be strongly affected by parity projection. This imaging technique allows us to capture higher-order interference peaks, which can be observed in the corners of the averaged image in Fig.~\ref{fig:fig5} (c). Having demonstrated the realization of a strontium quantum gas in the Bose-Hubbard regime, a natural next step will be the preparation of a unit-filled Mott insulator. 
This will be achieved by increasing the vertical confinement with a vertical optical lattice, which will soon allow us to reach the strongly interacting regime.

\section{\label{sec:conclusion}Conclusion and Outlook}

In this work we have demonstrated site-resolved imaging of \textsuperscript{84}Sr atoms in a square optical lattice. The imaging has been assisted by the realization of attractive Sisyphus cooling in an optical lattice. This technique has allowed us to realize a quantum-gas microscope with modest total laser powers ($\sim \SI{3}{W}$ at \SI{813.4}{nm}). In addition, we have probed the phase coherence in the system to demonstrate the realization of a strontium superfluid in the Hubbard regime. Since the entire atomic confinement is realized using the clock-magic wavelength of \SI{813.4}{nm}, our setup is prepared for spectroscopy measurements using the optical clock transition, enabling high-precision measurements of single and many-body features~\cite{franchi_statedependent_2017, bouganne_clock_2017, marti_imaging_2018, park_cavityenhanced_2022}. 

At its current stage, the experimental setup is already capable of realizing microscopic studies of dissipation in the Bose-Hubbard model~\cite{tomita_observation_2017, bouganne_anomalous_2020}. 
Another exciting research direction is that of collective light-matter phenomena in subwavelength atomic arrays~\cite{rui_subradiant_2020, hutson_observation_}, which we could realize by preparing a unit-filled Mott insulator. This is particularly exciting for strontium, since it exhibits micron-wavelength transitions that can be used to induce long-range dipolar interactions~\cite{olmos_longrange_2013}.  Finally, we stress that, while the results in this work are restricted to one bosonic isotope, our setup is capable of working with all stable isotopes of strontium~\cite{hoschele_atomnumber_2023}, and generalization of quantum-gas microscopy to the fermionic isotope \textsuperscript{87}Sr should be straightforward. This will open the door to microscopic studies of SU($N$) fermions, providing access to strongly interacting synthetic quantum Hall systems~\cite{zhou_observation_2023} and exotic forms of quantum magnetism~\cite{hermele_mott_2009, corboz_simultaneous_2011, taie_observation_2022}.\\

\begin{acknowledgments}
We acknowledge discussions with S. Blatt,  A. Kaufman, V. Klüsener, A. Park, P. Schau\ss,  A. Young, J. Zeiher, and the other members of the ICFO Quantum Gases Experimental group. We thank M. Miranda for the band-structure calculations, and S. Hirthe for contributions to the reconstruction algorithm and for a careful reading of the manuscript. 

We acknowledge funding from the European Union (PASQuanS2.1 project No. 101113690, DAALI project No. 899275, and ERC SuperComp project No. 101003295), MCIN/AEI/10.13039/501100011033 (LIGAS project PID2020-112687GB-C21, DYNAMITE QuantERA project PCI2022-132919 with funding from European Union NextGenerationEU, Equipamiento Científico Técnico EQC2018-005001-P, EQC2019-005699-P, and EQC2019-005706-P, Severo Ochoa CEX2019-000910-S, and PRTR-C17.I1 with funding from European Union NextGenerationEU and Generalitat de Catalunya), Fundació Cellex, Fundació Mir-Puig, and Generalitat de Catalunya (AGAUR 2021-SGR-01448 and CERCA program).
S.B. acknowledges support from MCIN/AEI/10.13039/501100011033 and ESF (PRE2020-094414), J.H. from the European Union (Marie Sk\l{}odowska-Curie–713729), V.M. from the Beatriu de Pinós Program and the Ministry of Research and Universities of the Government of Catalonia (2019-BP-00228), and A.R. from the MCIN/AEI/10.13039/501100011033 (Juan de la Cierva Formación FJC2020-043086-I).
\end{acknowledgments}

\appendix

\section{Details on the experimental setup\label{sec:appendixA}}

To prepare a cold cloud, prior to evaporative cooling, we capture atoms coming from an atomic source in the quartz glass cell (external dimensions of $67\, \times\,20\,\times\,$\SI{17}{mm}$^3$ with a glass thickness of \SI{3.5}{mm}) by means of a three dimensional magneto-optical trap (MOT) operated on the blue \SI{461}{nm} transition ($^1$S$_0$ to $^1$P$_1$) with a linewidth of $\Gamma/2 \pi = \, $\SI{30.5}{MHz}. We continuously drive the MOT with two repumper lasers at \SI{481}{nm}~\cite{hu_analyzing_2019} and \SI{679}{nm}~\cite{dinneen_cold_1999} to restore atoms leaking into the metastable states $^3 $P$_{J=0,\,2}$. The same repumper lasers are employed during fluorescence imaging, see Fig.~\ref{fig:fig1} (d). Additionally, we shield the atoms partially from losses of the blue MOT transition by resonantly driving the red \SI{689}{nm} transition, as described in our previous work~\cite{hoschele_atomnumber_2023}. In a second laser-cooling stage, we bring the atoms in a narrow-line red MOT to a temperature of few microkelvin, exploiting the \SI{7.4}{kHz} wide intercombination line at \SI{689}{nm} from $^1$S$_0$ to $^3$P$_1$  \cite{katori_magnetooptical_1999}. 
In order to maximize the transfer efficiency, we start the red MOT with frequency-broadened laser beams to address many velocity classes of the atoms~\cite{norcia_narrowline_2018, snigirev_fast_2019}. We then proceed by narrowing it to a single-frequency red MOT, resulting in around $5\times 10^5\,$atoms at temperatures of few microkelvin.
Then, we load them into a \SI{20}{\micro K} deep optical dipole trap at \SI{1064}{nm}, generated by a Nd:YAG laser (Mephisto MOPA 25W). This far-detuned crossed optical dipole trap consists of two beams with $\sim $\SI{100}{\micro m} waist at \SI{3}{W} initial power, propagating in plane and orthogonal to each other. We use it to load the atoms into the light-sheet trap as described in the main text.
For the light sheet and the lattice, which form the imaging potential, we use a Ti:sapphire laser (Matisse CS pumped by Millennia eV25W) at \SI{813.4}{nm}. Since both trap beams have the same vertical polarization, we avoid their interference by shifting their frequencies by few MHz with respect to each other.

\begin{figure}
\includegraphics[width= \columnwidth]{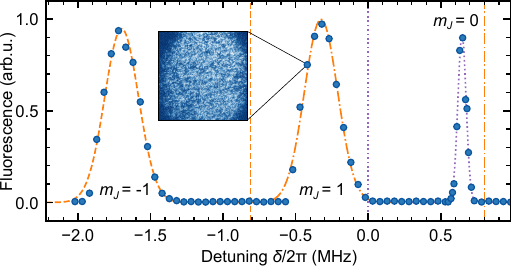}
\caption{Fluorescence spectrum of Sisyphus cooling. We detect the fluorescence of the atoms in the optical lattice (blue circles) for varying values of cooling frequencies detuned by $\delta$ from the free-space resonance of the $m_J = 0$ transition (dotted vertical line). The three peaks are fitted with three independent Gaussians, and correspond to the light-shifted $m_J$ transitions of the $^3 P _1$ excited state. The free-space resonance for the $|m_J| = 1$ transitions (dashed and dotted-dashed vertical lines) are centered around zero and Zeeman splitted by $\pm$\SI{0.8}{MHz}. Typically we detune the cooling light $\delta/2\pi \approx \SI{-0.4}{MHz}$ from the free-space resonance in order to cool a large region, see inset.}
\label{fig:fig6}
\end{figure}

\section{Optical lattice\label{sec:appendixB}}

The bow-tie configuration of the optical-lattice beams results in a four-fold interference, as depicted in Fig.~\ref{fig:fig1} (a). Thus, the lattice potential can be written as
\begin{equation}
    \begin{split}
         V_{\text{lat}}(x, y) = -\frac{V_0}{4}  \{ \cos^2 (k x) &+ \cos^2(k y) \\
         &+  2 \cos(k x) \cos(k y) \} \text{,}
         \label{eq:eq1}
    \end{split}
\end{equation}
where $k = 2 \pi / \lambda $ describes the wavevector with $\lambda = \SI{813.4}{nm}$. $V_0$ is the full lattice depth, which displays a four-fold enhancement with respect to the single retroreflected-beam case. Moreover, the lattice spacing is $\lambda/\sqrt{2} \approx  \SI{575}{nm}$ and not $\lambda/2 \approx \SI{407}{nm}$. In this work, we express $V_0$ in terms of the recoil energy $E_r = h^2/2 m \lambda^2 \approx \, h \times \SI{3.6}{kHz}$ of the lattice photons, and not of the characteristic energy scale fixed by the lattice spacing $h^2/8 m a^2 \approx \, h \times \SI{1.8}{kHz}$. This is a convenient choice when considering alternative lattice geometries~\cite{sebby-strabley_lattice_2006, tarruell_creating_2012, wei_observation_2023}, which we plan to use in future experiments. 
Additionally, this definition is useful in the presence of small variations of the lattice spacing along $x$ and $y$ directions, as those observed in our experiment ($a_x=\SI{608(1)}{nm}$ and $a_y=\SI{547(1)}{nm}$). This spacing anisotropy stems from a small deviation of the lattice beams from the perpendicular orientation.
Considering the relation between the lattice spacings and the angle $\alpha_{x,y}$ between the lattice beams
\begin{equation*}
    a_{x,y} = \frac{\lambda}{2} \frac{1}{\sin \left( \alpha_{x,y} / 2 \right)},
\end{equation*}
we find a $6^{\circ}$ angle deviation.
Another modification from the ideal lattice potential given in Eq. \ref{eq:eq1} comes from the uncoated glass cell, which reduces the power of the lattice beams ($\approx \,$\SI{86}{\%} transmission through each wall of the cell).
By incorporating all these effects into a full band-structure calculation we obtain an interaction strength $U$ that differs by less than $3\,\%$ from the ideal model, and  anisotropic tunnelings of $J_x/h = \,$\SI{161(7)}{Hz} and $J_y/h = \,$\SI{222(8)}{Hz}. These parameters still correspond to the superfluid Hubbard regime, as discussed in the main text.\\

\section{Spectrum of red cooling compared to free space\label{sec:appendixC}}
To gain understanding of the cooling spectrum during site-resolved imaging, we compare the cooling frequencies with the free-space resonances at the magnetic field configuration ($\vec{B} \parallel \vec{\epsilon}_{813}$) that we use, see main text. The free-space resonance is measured by shining resonant red light (\SI{689}{nm}) in absence of any optical dipole trap potential and then determining the surviving atoms in absorption imaging.

The three resonances we measure correspond to the three $m_J$ states of the $^3 P _1$ excited state, displayed by the vertical lines in Fig.~\ref{fig:fig6}. The $m_J=\pm 1$ are Zeeman splitted from the $m_J = 0$ resonance by $\pm$\SI{0.8}{MHz} and allow to determine the magnitude of the magnetic field $B_z\approx$ \SI{0.58}{G}.

The spectrum of the cooling transition is inferred from the site-resolved fluorescence images, because atoms are only cooled efficiently with attractive Sisyphus cooling if the red cooling light is on or close to resonance. The measurement is performed along the same lines as the one of Fig.~\ref{fig:fig4}. The three resonances at which we detect single atoms thanks to cooling are light shifted due to the lattice and light sheet potential. In this work we cool on the $m_J = 1$ transition, however the $m_J = -1$ works equivalently, see Fig. \ref{fig:fig6}. The third resonance corresponding to $m_J = 0$ is less effective in cooling. For this state, we have repulsive Sisyphus cooling due to its different polarizability compared to the $|m_J| = 1$ states.

We observe the maximum fluorescence signal of the $m_J = 1$ state \SI{1.1(1)}{MHz} shifted from its free-space resonance (dotted-dashed orange vertical line in Fig.~\ref{fig:fig6}).
While we typically perform cooling around this detuning, the light-shifted resonance at the center of the trap, lies at the lower edge of the Gaussian fit, namely at $\delta \approx -2 \pi \times \SI{0.5}{MHz}$. This corresponds to a \SI{1.3}{MHz} detuning from its free-space resonance.

\end{document}